# Social Networking Site For Self Portfolio

N. Sampath Kumar[1], U. KarthikChandran[2], N.ArunKumar[3], K. Karnavel[4]

[1,2,3] UG Student, Department of Computer Science & Engg,
Anand Institute of Higher Technology, Chennai – 603 103

[4] Assistant Professor, Department of Computer Science & Engg,
Anand Institute of Higher Technology, Chennai-603 103

**Abstract**

Online social networking concept is a global phenomenon and there are millions of sites which help in being connected with friends and family. This project focuses on creating self-portfolios for the users which makes the users engaging with their skills. The users follow the other users to interact and communicate with them. Users can encourage the other user's blogs and videos by clicking the hit button.The functionality of this site is designed to focus on both professional as well as academics. Each user is given a dashboard for uploading videos and writing blogs.

***Keywords*:** *Network Visualization Social network analysis Information categorization Information analysis*

## 1. Introduction

Since their introduction, social network sites (SNSs) such as MySpace, Facebook, Cyworld, and Bebo have attracted millions of users, many of whom have integrated these sites into their daily practices. As of this writing, there are hundreds of SNSs, with various technological affordances,supporting a wide range of interests and practices. While their key technological features are fairly consistent, the cultures that emerge around SNSs are varied. Most sites support the maintenance of pre-existing social networks, but others help strangers connect based on shared interests, political views, or activities. Some sites cater to diverse audiences, while others attract people based on common language or shared racial, sexual, religious, or nationality-based identities. Sites also vary in the extent to which they incorporate new information and communication tools, such as mobile connectivity, blogging, and photo/video-sharing.

Scholars from disparate fields have examined SNSs in order to understand the practices, implications, culture, and meaning of the sites, as well as users' engagement with them. This special theme section of the Journal of Computer-Mediated Communication brings together aunique collectionof articles that analyze a wide spectrum of social network sites using various methodological techniques, theoretical traditions, and analytic approaches. By collecting these articles in this issue, our goal is to showcase some of the interdisciplinary scholarship around these sites[7].

The purpose of this introduction is to provide a conceptual, historical, and scholarly context for the articles in this collection. We begin by defining what constitutes a social network site and then present one perspective on the historical development of SNSs, drawing from personal interviews and public accounts of sites and their changes over time. Following this, we review recent scholarship on SNSs and attempt to contextualize and highlight key works. We conclude with a description of the articles included in this special section and suggestions for future research.

## 2. Portfolio

Rick Stiggins (1994) defines a portfolio as a collection of student work that demonstrates achievement or improvement. The material to be collected and the story to betold can vary greatly as a function of the assessment context.

The Northwest Evaluation Association offers a similar definition: A purposeful collection of student work that illustrates efforts, progress, and achievement in one or moreareas [over time]. The collection must include: student participation in selecting contents, the criteria forselection, the criteria for judging merit, and evidence of student self-reflection.

## 3. Web service

A **Web service** is a method of communication between two electronic devices over the World Wide Web. A **Web service** is a software function provided at a network address over the web or the cloud, it is a service that is "always on" as in the concept of utility computing [3].

### 3.1 Motivation of research problem

1. What are the benefits of developing self-portfolios in social networking site?
2. What are the benefits of uploading certificates in our profile?
3. What are the benefits of using chat rooms in our site?

## 4. Social Network Sites

We define social network sites as web-based services that allow individuals to (1) construct a public or semi-public profile within a bounded system, (2) articulate a list of other users with whom they share a connection, and (3) view and traverse their list of connections and those made by others within the system. The nature and nomenclature of these connections may vary from site to site.

While we use the term "social network site" to describe this phenomenon, the term "social networking sites" also appears in public discourse, and the two terms are often used interchangeably. We chose not to employ the term "networking" for two reasons: emphasis and scope. "Networking" emphasizes relationship initiation, often between strangers. While networking is possible on these sites, it is not the primary practice on many of them, nor is it what differentiates them from other forms of computer-mediated communication (CMC) [5].

What makes social network sites unique is not that they allow individuals to meet strangers, but rather that they enable users to articulate and make visible their social networks. This can result in connections between individuals that would not otherwise be made, but that is often not the goal, and these meetings are frequently between "latent ties" (Haythornthwaite, 2005) who share some offline connection. On many of the large SNSs, participants are not necessarily "networking" or looking to meet new people; instead, they are primarily communicating with people who are already a part of their extended social network. To emphasize this articulated social network as a critical organizing feature of these sites, we label them "social network sites."

### 4.1 A History of Social Network Sites The Early Years

According to the definition above, the first recognizable social network site launched in 1997. SixDegrees.com allowed users to create profiles, list their Friends and, beginning in 1998, surf the Friends lists. Each of these features existed in some form before SixDegrees, of course. Profiles existed on most major dating sites and manycommunity sites. AIM and ICQ buddy lists supported lists of Friends, although those Friends were not visible to others. Classmates.com allowed people to affiliate with their high school or college and surf the network for others who were also affiliated, but users could not create profiles or list Friends until years later. SixDegrees was the first to combine these features.

SixDegrees promoted itself as a tool to help people connect with and send messages to others. While SixDegrees attracted millions of users, it failed to become a sustainable business and, in 2000, the service closed. Looking back, its founder believes that SixDegrees was simply ahead of its time (A. Weinreich, personal communication, July 11, 2007). While people were already flocking to the Internet, most did not have extended networks of friends who were online. Early adopters complained that there was little to do after accepting Friend requests, and most users were not interested in meeting strangers[10].

From 1997 to 2001, a number of community tools began supporting various combinations of profiles and publicly articulated Friends. AsianAvenue, BlackPlanet, and MiGente allowed users to create personal, professional, and dating profiles—users could identify Friends on their personal profiles without seeking approval for those connections (O. Wasow, personal communication, August 16, 2007). Likewise, shortly after its launch in 1999, LiveJournal listed one-directional connections on user pages. LiveJournal's creator suspects that he fashioned these Friends after instant messaging buddy lists (B. Fitzpatrick, personal communication, June 15, 2007)—on LiveJournal, people mark others as Friends to follow their journals and manage privacy settings. The Korean virtual worlds site Cyworld was started in 1999 and added SNS features in 2001, independent of these other sites (see Kim & Yun, this issue). Likewise, when the Swedish web community LunarStorm refashioned itself as an SNS in 2000, it contained Friends lists, guestbooks, and diary pages (D. Skog, personal communication, September 24, 2007).

The next wave of SNSs began when Ryze.com was launched in 2001 to help people leverage their business networks. Ryze's founder reports that he first introduced the site to his friends—primarily members of the San Francisco business and technology community, including the entrepreneurs and investors behind many future SNSs (A. Scott, personal communication, June 14, 2007). In particular, the people behind Ryze, Tribe.net, LinkedIn, and Friendster were tightly entwined personally and professionally. They believed that they could support each other without competing (Festa, 2003). In the end, Ryze never acquired mass popularity, Tribe.net grew to attract a passionate niche user base, LinkedIn became a powerful business service, and Friendster became the most significant, if only as "one of the biggest disappointments in Internet history" (Chafkin, 2007, p. 1)[12].

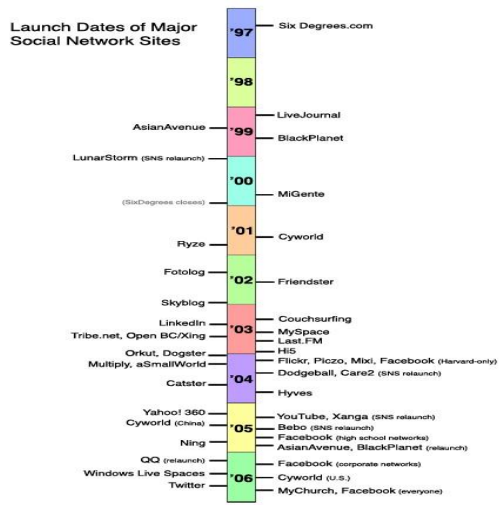

Figure 1. Timeline of the launch dates of many major SNSs and dates when community sites re-launched with SNS features. Like any brief history of a major phenomenon, ours is necessarily incomplete. In the following section we discuss Friendster, MySpace, and Facebook, three key SNSs that shaped the business, cultural, and research landscape.

4.2 Networks and Network Structure

Social network sites also provide rich sources of naturalistic behavioral data. Profile and linkage data from SNSs can be gathered either through the use of automated collection techniques or through datasets provided directly from the company, enabling network analysis researchers to explore large-scale patterns of friending, usage, and other visible indicators (Hogan, in press), and continuing an analysis trend that started with examinations of blogs and other websites [12]. For instance, Golder, Wilkinson, and Huberman (2007) examined an anonymized dataset consisting of 362 million messages exchanged by over four million Facebook users for insight into Friending and messaging activities. Lampe, Ellison, and Steinfield (2007) explored the relationship between profile elements and number of Facebook friends, finding that profile fields that reduce transaction costs and are harder to falsify are most likely to be associated with larger number of friendship links. These kinds of data also lend themselves well to analysis through network visualization (Adamic, Büyükkökten, & Adar, 2003; Heer&boyd, 2005; Paolillo& Wright, 2005).

## 5. Research objective

1. The most important benefit of developing self-portfolio in social networking site is solving storage problem associated with traditional paper based portfolios,additionally instructors can easily comment on learners work by tap into the portfolio,also anyone in the world can be granted access to the portfolio and students can have perfect control on what artifacts can be presented and who can see them.

2. The important benefit of uploading certificates in our profile is solving the problem of taking the traditional paper based certificates for an interview; attending an online examination these certificates in our profile can be used.

3. Chat rooms are available for the users who wants to interact and communicate with the professionals in any domain the user is interested discuss about the ideas, ask solutions about the problems.

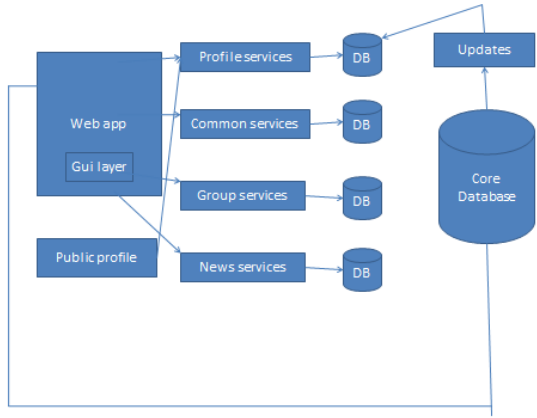

## 6. Conclusion

Portfolio can facilitate long-life, meaningful learning. Besides the students developed more self-monitoring and self-regulation and self-assessment by producing more reflective works. In the recent mode of instruction the educators can keep in touch with learners and provide editing fast online feedback as a guide and simulator and facilitator rather than proving the correct answer. Taking participants reflections into consideration the teacher can promote learners responsibility and feeling of ownership toward their portfolio. Consequently it would support students to use portfolio as a tool in order to promote their learning process.